\documentstyle[12pt,axodraw,epsf,epsfig]{article}

\newcommand{\beq}{ \begin{eqnarray} }
\newcommand{\eeq}{ \end{eqnarray} }
\newcommand{\beqstar}{ \begin{eqnarray*} }
\newcommand{\eeqstar}{ \end{eqnarray*} }

\newcommand{\lsim}{ \mathop{}_{\textstyle \sim}^{\textstyle <} }

\begin{document}
\baselineskip 0.7cm

\begin{titlepage}

\begin{center}

\hfill ICRR-Report-505-2004-3\\
\hfill TU-724\\
\hfill \today

{\large 
Hadronic EDMs in SUSY SU(5) GUTs with Right-handed Neutrinos
}
\vspace{1cm}

{\bf Junji Hisano}$^{1}$,  
{\bf Mitsuru Kakizaki}$^{1}$,
{\bf Minoru Nagai}$^{1}$
and 
{\bf Yasuhiro Shimizu}$^{2}$
\vskip 0.15in
{\it
$^1${ICRR, University of Tokyo, Kashiwa 277-8582, Japan }\\
$^2${Department of Physics, Tohoku University, Sendai 980-8578, Japan}\\
}
\vskip 0.5in

\abstract{
We discuss hadronic EDM constraints on the neutrino sector in the SUSY
SU(5) GUT with the right-handed neutrinos.  The hadronic EDMs are
sensitive to the right-handed down-type squark mixings, especially
between the second and third generations and between the first and
third ones, compared with the other low-energy hadronic observables,
and the flavor mixings are induced by the neutrino Yukawa
interaction. The current experimental bound of the neutron EDM may
imply that the right-handed tau neutrino mass is smaller than about
$10^{14}$ GeV in the minimal supergravity scenario, and it may be
improved furthermore in future experiments, such as the deuteron EDM
measurement. }
\end{center}
\end{titlepage}
\setcounter{footnote}{0}

The supersymmetric grand unified models (SUSY GUTs) are  ones of
the well-motivated models after discovery of the gauge coupling
unification at the LEP experiment. Non-vanishing light neutrino
masses shown in the neutrino oscillation experiments might also
suggest existence of the SUSY GUTs since the right-handed neutrino
masses expected from the measurements are near the GUT scale in the
seesaw mechanism \cite{seesaw}.  Nowadays many efforts are devoted to
search for the next signature from both theoretical and experimental
sides.

It is well-known that the SUSY GUTs predict rich flavor violation in
the SUSY breaking terms for the squarks and sleptons when origin of
the SUSY breaking is the dynamics above the GUT scale, such as the
gravity mediation \cite{Hall:1985dx,bh}.  In the minimal SUSY standard
model (MSSM), the sizable flavor-violating SUSY breaking terms for the
left-handed squarks are induced by the large top-quark Yukawa
coupling, and those for the left-handed sleptons may be also generated
by the neutrino Yukawa interaction in the SUSY seesaw mechanism
\cite{bm,lfv}. In the SUSY GUTs the flavor-violating SUSY breaking terms
for the SU(5) partners of the left-handed squarks or sleptons,
right-handed squarks and sleptons, are generated by the
flavor-violating interactions for the colored Higgs multiplets, that 
are also SU(5) partners of the doublet Higgs multiplets. This
gives a chance to probe the interactions at the GUT scale by the
low-energy flavor-changing processes, such as $K^0$--$\overline{K}^0$
mixing, $B$ physics, and lepton flavor violation.

On the other hand, the colored Higgs interactions may have the
CP-violating phases, which are independent of the CKM phase in the
MSSM. The CP-violating phases may contribute to the electric dipole
moments (EDMs) of electron, neutron and atoms
\cite{Dimopoulos:1994gj,Barbieri:1995tw,Khriplovich:1996gk,Romanino:1996cn} via the 
flavor-violating SUSY breaking terms. While the EDMs are
flavor-conserving processes, they may depend on the flavor-violating
SUSY breaking terms in the squark or slepton internal lines. Moreover,
since both the left-handed and right-handed squarks or sleptons have
the flavor-violating SUSY breaking mass terms in the SUSY GUTs, the
EDMs are enhanced by the heavier fermion masses. Thus, the EDMs are
good probes for the SUSY GUTs.

In this paper we discuss the hadronic EDMs in the SUSY SU(5) GUT with
the right-handed neutrinos. The neutrino Yukawa coupling generates the
flavor-violating SUSY breaking terms for the right-handed down-type
squarks, which are SU(5) partners of the left-handed sleptons. This
implies that we can investigate the neutrino sector by the hadronic
EDMs. We show that the structure in the neutrino sector is more
constrained by the current experimental bounds for the hadronic EDMs
than by the other hadronic observables. As shown in
Ref.~\cite{Khriplovich:1996gk,HS,HS2}, the hadronic EDMs depend on the
chromoelectric dipole moment (CEDM) for the strange quark in addition
to those for the up and down quarks. It implies that the hadronic EDMs
give the strong constraints on the mixings for the right-handed
down-type squarks between the first and  third and between the
second and third generations.

A new technique for measurement of the deuteron EDM is proposed recently
\cite{Semertzidis:2003iq}, and the sensitivity  may reach to $d_D
\sim(1-3)\times 10^{-27}e\,cm$. We show the sensitivity of
the experiment to the SUSY SU(5) GUT with the right-handed neutrinos.
We also discuss dependence of the prediction for the hadronic EDMs on
the SUSY breaking models.

First, we review the flavor structure in the squark and slepton mass
matrices in the SUSY SU(5) GUT with the right-handed neutrinos. The
Yukawa interactions for quarks and leptons and the Majorana mass terms
for the right-handed neutrinos in this model are given by the following
superpotential,
\begin{eqnarray}
W&=& 
\frac14 f_{ij}^{u} \Psi_i \Psi_j H 
+\sqrt{2} f_{ij}^{d} \Psi_i \Phi_j \overline{H}
+f_{ij}^{\nu} \Phi_i \overline{N}_j {H}
+M_{ij} \overline{N}_i \overline{N}_j,
\label{superp_gut}
\end{eqnarray}
where $\Psi$ and $\Phi$ are for {\bf 10}- and {$\bf \bar{5}$}-dimensional
multiplets, respectively, and $\overline{N}$ is for the right-handed
neutrinos.  $H$ ($\overline{H}$) is {\bf 5}- ({$\bf \bar{5}$}-)
dimensional Higgs multiplets.  After removing the unphysical degrees
of freedom, the Yukawa coupling constants in Eq.~(\ref{superp_gut})
are given as follows,
\begin{eqnarray}
f^u_{ij} &=& 
V_{ki} f_{u_k} {\rm e}^{i \varphi_{u_k}}V_{kj}, \nonumber\\
f^d_{ij} &=& f_{d_i} \delta_{ij},\nonumber\\
f^\nu_{ij} &=& {\rm e}^{i \varphi_{d_i}} 
U^\star_{ij} f_{\nu_j}.
\label{yukawa}
\end{eqnarray}
Here, $\varphi_{u}$ and $\varphi_{d}$ are CP-violating phases inherent
in the SUSY SU(5) GUT. They satisfy $\sum_i \varphi_{f_i} =0$
$(f=u$ and $d)$.  The unitary matrix $V$ is the CKM matrix in the extension
of the SM to the SUSY SU(5) GUT, and each unitary matrices $U$ and $V$
have only a phase. When the Majorana mass matrix for the right-handed
neutrinos is diagonal in the basis of Eq.~(\ref{yukawa}), $U$ is the MNS matrix
observed in the neutrino oscillation.  In this paper we assume the
diagonal Majorana mass matrix in order to avoid the complexity due to
the structure. In this case the light neutrino mass eigenvalues are
given as
\begin{eqnarray}
m_{\nu_{i}}
&=&
\frac{f_{\nu_i}^2}{M_{N_i}} \langle H_f \rangle^2
\label{seesaweq}
\end{eqnarray}
where $H_f$ is a doublet Higgs in $H$. 

The colored Higgs multiplets $H_c$ and $\overline{H}_c$ are introduced
in $H$ and $\overline{H}$ as SU(5) partners of the Higgs doublets in
the MSSM, respectively. They have new flavor-violating interactions in
Eq.~(\ref{superp_gut}). If the SUSY-breaking terms in the MSSM are
generated by dynamics above the colored Higgs masses, such as in
the gravity mediation, the sfermion mass terms may get sizable
corrections by the colored Higgs interactions. The interactions are
also baryon-number violating, and then proton decay induced by the
colored Higgs exchange is a serious problem, especially in the minimal
SUSY SU(5) GUT \cite{dim5}. However, it depends on the detailed
structure in the Higgs sector \cite{dim5sup,dim5sup6}. Thus, we ignore
the proton decay while we adopt the minimal Yukawa structure in
Eq.~(\ref{superp_gut}).

In the minimal supergravity scenario the SUSY breaking terms are
supposed to be given at the reduced Planck mass scale ($M_G$). In this
case, the flavor-violating SUSY breaking mass terms at low energy are
induced by the radiative correction, and they  are qualitatively given
in a flavor basis as
\begin{eqnarray}
(m_{{\tilde{u}_L}}^2)_{ij}  &\simeq&-
V_{i3}V_{j3}^\star  \frac{f_{b}^2}{(4\pi)^2}
\;\; (3m_0^2+ A_0^2) \;\; 
(2 \log\frac{M_G^2}{M_{H_c}^2}+ \log\frac{M_{H_c}^2}{M_{SUSY^2}}),\nonumber\\
(m_{\tilde{u}_R}^2)_{ij}  &\simeq& -
{\rm e}^{-i\varphi_{u_{ij}}} 
V_{i3}^\star V_{j3} \frac{2f_{b}^2}{(4\pi)^2}
\;\; (3m_0^2+ A_0^2) \;\; 
\log\frac{M_G^2}{M_{H_c}^2}, \nonumber\\
(m_{{\tilde{d}_L}}^2)_{ij}  &\simeq&-
V_{3i}^\star
V_{3j} \frac{f_{t}^2}{(4\pi)^2} 
\;\; (3m_0^2+ A_0^2) \;\; 
(3 \log\frac{M_G^2}{M_{H_c}^2}+ \log\frac{M_{H_c}^2}{M_{SUSY}^2}),\nonumber\\
(m_{\tilde{d}_R}^2)_{ij}  &\simeq&-
{\rm e}^{i\varphi_{d_{ij}}}  U^\star_{ik}U_{jk} 
\frac{f_{\nu_k}^2}{(4\pi)^2} 
\;\; (3m_0^2+A_0^2) \;\; 
\log\frac{M_G^2}{M_{H_c}^2},\nonumber\\
(m_{\tilde{l}_L}^2)_{ij}  &\simeq&-
U_{ik}U_{jk}^\star
\frac{f^2_{\nu_k} }{(4\pi)^2} 
\;\; (3m_0^2+ A_0^2) \;\; 
\log\frac{M_G^2}{M_{N_k}^2},\nonumber\\
(m_{{\tilde{e}_R}}^2)_{ij}  &\simeq&-
{\rm e}^{i\varphi_{d_{ij}}} 
V_{3i}V^\star_{3j}
\frac{3 f_{t}^2}{(4\pi)^2} 
\;\; (3m_0^2+ A_0^2)\;\; 
\log\frac{M_G^2}{M_{H_c}^2},
\label{sfermionmass}
\end{eqnarray}
with $i\ne j$, where
$\varphi_{u_{ij}}\equiv\varphi_{u_{i}}-\varphi_{u_{j}}$ and
$\varphi_{d_{ij}}\equiv\varphi_{d_{i}}-\varphi_{d_{i}}$ and
$M_{H_c}$ is the colored Higgs mass.  Here,
$M_{SUSY}$, $m_0$ and $A_0$ are the SUSY-breaking scale in the MSSM
and the universal scalar mass and trilinear coupling,
respectively. $f_t$ is the top quark Yukawa coupling constant  while $f_b$ is
for the bottom quark. As mentioned above, the off-diagonal components in the
right-handed squarks and slepton mass matrices are induced by the
colored Higgs interactions, and they depend on the CP-violating phases
in the SUSY SU(5) GUT with the right-handed neutrinos
\cite{Moroi:2000tk}.

When both the left-handed and right-handed squarks have the
off-diagonal components in the mass matrices, the EDMs and CEDMs for
the light quarks are enhanced significantly by the heavier quark mass.
The CEDMs are generated by the diagram in Fig.~1. In the SUSY SU(5)
GUT with the right-handed neutrinos, the neutrino Yukawa coupling
induces the flavor-violating mass terms for the right-handed down-type
squarks.  Since the flavor-violating mass terms for the left-handed
down-type squarks are expected to be dominated by the radiative
correction induced by the top quark Yukawa coupling as in
Eq.~(\ref{sfermionmass}), we can
investigate or constrain the structure in the neutrino sector.

The CEDMs for the light quarks, including the strange quark,
contribute to the hadronic EDMs since the CP-violating nucleon
coupling is induced. Here, we consider only the CEDMs for the light
quarks.\footnote{ In Ref.~\cite{Khriplovich:1996gk}, the hadronic EDMs
are discussed in the SUSY SO(10) GUT, including down and strange
CEDMs.}  While the EDMs for the up and down quarks contribute to the
neutron EDM, we find that the EDM contributions are a factor smaller
than the CEDM contributions. In this evaluation we use the prediction
of non-relativistic quark model for the light quark EDM contribution
to neutron EDM and the estimation by the chiral Lagrangian for the
CEDM contribution \cite{HS2}.  While the ratio may suffer from the
theoretical uncertainties, it is safe to ignore the light quark EDM
contribution for the purpose of the demonstration of the hadronic EDM
sensitivity.  The EDM of ${^{199}}$Hg atom, which is not sensitive to
the quark EDMs, also gives similar bounds on the CEDMs to the
neutron EDM as shown later.

The CEDMs of the light quarks derived by the flavor violation in the
both the left-handed and right-handed quark mass matrices are given by
the following dominant contribution, which is enhanced by the heavier
quark masses,\footnote{ The CEDM of the up quark is also generated by
diagrams with single and double mass insertions in the SUSY SU(5) GUT
with the right-handed neutrinos.  They may be comparable to those with
triple mass insertions in Eq.~(\ref{SUSYEDM}), since $A^{(u)}$ may
have flavor violation. In this paper we concentrate on the CEDMs for
the down and strange quarks, thus, we ignore them in this paper.  }
\begin{eqnarray}
 {d}^C_{q_i} = c\frac{\alpha_s}{4\pi}\frac{m_{\tilde g}}{\overline{m}^2_{\tilde q}}
f\left(\frac{m_{\tilde{g}}^2}{\overline{m}^2_{\tilde{q}}}
\right)  {\mathrm Im}
\left[(\delta^q_{ij})_{L}(\delta^q_{j})_{LR}(\delta^q_{ji})_{R}
\right],
\label{SUSYEDM}
\end{eqnarray}
where $m_{\tilde{g}}$ and $\overline{m}_{\tilde{q}}$ are the gluino and averaged
squark masses and $c$ is the QCD correction, $c\sim 0.9$.  The mass
insertion parameters are defined as
\begin{eqnarray}
(\delta^f_{ij})_{L/R}&\equiv&
\frac{(m_{{\tilde{f}_{L/R}}}^2)_{ij}}{(\overline{m}_{{\tilde{f}}}^2)},
\nonumber\\
( \delta_{i}^{d})_{LR} 
&\equiv&  \frac{m_{d_i}(A_i^{(d)} -\mu\tan\beta)}{\overline{m}^2_{\tilde{d}}},
\nonumber\\
( \delta_{i}^{u})_{LR} 
&\equiv& \frac{m_{u_i}(A_i^{(u)} -\mu\cot\beta)}{\overline{m}^2_{\tilde{u}}}.
\end{eqnarray}
The function $f(x)$ is given as
\begin{eqnarray}
f(x)&=&\frac{177+118x-288x^2-6x^3-x^4+(54+300x+126x^2)\log x}{18(1-x)^6},
\label{mass_func}
\end{eqnarray}
and $f(1) =-11/180$.\footnote{
We find the numerical errors in the expression of the QCD correction $c$
and the mass function $f(x)$ given in Ref.~\cite{HS}.
}

In order to translate the CEDMs of the light quarks to the EDMs of 
neutron and ${^{199}}$Hg atom,  we use the evaluation of 
the  EDMs of neutron
and ${^{199}}$Hg atom in Ref.~\cite{HS2} as\footnote{
The hadronic EDMs, especially neutron EDM, depend on whether the
Peccei-Quinn symmetry works or not \cite{Bigi}, though the numerical
difference is small. We assume the PQ symmetry in this paper. When
both the left-handed and right-handed squarks have off-diagonal terms
in the mass matrices, the one-loop correction to the light quark
masses by the SUSY loops may induce the QCD theta term even in a case
where the tree-level QCD theta parameter is zero.  Thus, this
assumption is natural.
}
\begin{eqnarray}
d_n &=& -1.6 \times e (d_u^C+0.81\times d_d^C+0.16\times d_s^C),
\nonumber\\
d_{\rm Hg}&=&-8.7\times 10^{-3}\times e(d_u^C-d_d^C+0.005 d_s^C),
\end{eqnarray}
where $d_n$ is generated by the charged meson loops and $d_{\rm Hg}$
comes from the nuclear force by the pion exchange
in the chiral perturbation theory.  The
experimental upperbounds on the EDMs of neutron \cite{Harris:jx} and
$^{199}$Hg atom \cite{Romalis:2000mg} are
\begin{eqnarray}
 |d_{n}| < 6.3 \times 10^{-26} e\, cm,
\nonumber\\
 |d_{\rm Hg}| < 1.9\times 10^{-28} e\, cm,
\label{dhbound}
\end{eqnarray}
respectively (90\%C.L.). Thus, the upperbounds on the quark CEDMs are
\begin{eqnarray}
e|{d}^C_u|&<&3.9(2.2)\times 10^{-26}\;e\;cm, \nonumber\\
e|{d}^C_d|&<&4.8(2.2)\times 10^{-26}\;e\;cm,\nonumber\\
e|{d}^C_s|&<&2.4(44) \times 10^{-25}\;e\;cm,
\label{nedm_pq}
\end{eqnarray}
from the EDM of neutron ($^{199}$Hg atom). Here, we assume that the
accidental cancellation among the CEDMs does not suppress the EDMs.
The constraint on ${d}^C_s$ from $^{199}$Hg atom is one-order weaker
than that from neutron, since the contribution to the EDM of $^{199}$Hg
atom is suppressed by $\pi^0$-$\eta^0$ mixing \cite{HS2}.

Now we show the significance of the hadronic EDMs for investigation of
the flavor-violating terms in the squark mass matrices. In Table~1 we
show the current constraints on the mass insertion parameters for
squarks from various hadronic processes including the EDMs of neutron
and $^{199}$Hg atom. For completeness, we include the constraints on
the mass insertion parameters for the up-type squarks. The bounds are
derived from formulas in Refs.~\cite{Gabbiani:1996hi,Hisano:1998fj}.
For the CP-violating observables, we assume that the related
CP-violating phases in the squark mass matrices are $O(1)$. In the
MSSM the mass insertion parameters for the left-handed down-type
squarks are typically
\begin{eqnarray}
(\delta^d_{12})_{L}\sim\lambda^5,\,\,~~
(\delta^d_{13})_{L}\sim\lambda^3,\,\,~~
(\delta^d_{23})_{L}\sim\lambda^2,
\end{eqnarray}
with $\lambda\sim 0.2$, since they are generated by the top quark
Yukawa coupling. Combined with these relations, we find that
$(\delta_{12}^d)_R \lsim 10^{-4}$ from $\epsilon_K$,
$(\delta_{13}^d)_R \lsim 10^{-(2-3)}$ from the EDMs of neutron and
$^{199}$Hg atom, $(\delta_{23}^d)_R \lsim 10^{-(2-3)}$ from the
neutron EDM in the case of $m_{SUSY}=500$GeV and $\tan\beta=10$. The
other processes are less significant compared with $\epsilon_K$ and
the hadronic EDMs.  While $\epsilon_K$ gives a constraint on a product
$(\delta_{13}^d)_R (\delta_{32}^d)_R$, the EDM constraints are
severer, especially, in large $\tan\beta$.  This means that we can
probe efficiently the flavor violation of the tau neutrino Yukawa
coupling by the hadronic EDMs, since the Yukawa coupling generates the
non-vanishing $(\delta_{13}^d)_R$ and $(\delta_{23}^d)_R$.

It should be careful to compare the EDM constraints with other
low-energy observables, which are theoretically controlled better,
since the hadronic EDMs may suffer from more hadronic
uncertainties. However, the orders of the magnitude in the EDM
constraints are still expected to have the significance. Recently, the Belle
collaboration announced that CP asymmetry in $B\to
\phi K_s$ is $-0.96\pm 0.50$, which is 3.5 $\sigma$ deviation from the
SM prediction \cite{Abe:2003yt}, while the Babar result on the CP
asymmetry is consistent with the SM prediction
\cite{babar}.  The deviation may be explained by introduction of
the right-handed bottom and strange squarks mixing, such as in the
SUSY SU(5) GUT with the right-handed neutrinos, however, the neutron
EDM constraint on the strange quark CEDM gives an upperbound on the
deviation \cite{HS2}. Even if the discrepancy comes from the hadronic
uncertainty in the evaluation of the neutron EDM, the further
improvement of the bound on the hadronic EDMs is very important.

The neutrino Yukawa interaction induces the flavor-violating mass
terms for the left-handed sleptons, which are related to those for the
right-handed down-type squarks \cite{Hisano:2003bd},
\begin{eqnarray}
(\delta_{ij}^d)_R&\simeq&
\frac{(\overline{m}_{{\tilde{l}}}^2)}{(\overline{m}_{{\tilde{d}}}^2)}
(\delta_{ij}^l)_L^\star,
\label{gutrelation}
\end{eqnarray}
in the SUSY SU(5) GUT with the right-handed neutrinos. Here,
$(\delta^l_{ij})_{L}\equiv
(m_{{\tilde{l}_{L}}}^2)_{ij}/(\overline{m}_{{\tilde{l}}}^2)$.  In
Table~2, we show the constraints on the mass insertion parameters for
sleptons from the lepton-flavor violating decay of the charged leptons
and the electron EDM. In the table, we take $m_{SUSY}=200$GeV and
$\tan\beta=10$.  The current bound on $Br(\mu\rightarrow e\gamma)$
gives a stringent constraint on $(m_{{\tilde{l}_{L}}}^2)_{12}$. On the
other hand, those on $(m_{{\tilde{l}_{L}}}^2)_{13}$ and
$(m_{{\tilde{l}_{L}}}^2)_{23}$ from the tau LFV decay are weaker than
the hadronic EDM constraints under the assumption of
Eq.~(\ref{gutrelation}).  The hadronic EDM constraints imply
$Br(\tau\rightarrow \mu (e)\gamma)\lsim 10^{-(7-8)}$.  While
$Br(\mu\rightarrow e\gamma)$ can give a constraint on them, it is a
product $(m_{{\tilde{l}_{L}}}^2)_{13}(m_{{\tilde{l}_{L}}}^2)_{32}$.
The electron EDM may give a constraint on a product
$(m_{{\tilde{e}_{R}}}^2)_{13}(m_{{\tilde{l}_{L}}}^2)_{31}$ while
$(m_{{\tilde{e}_{R}}}^2)_{13}$ depends on the CKM mixing at the GUT
scale. Thus, the hadronic EDMs are more directly related to the
neutrino Yukawa coupling when the CP violating phases are maximal.

In Fig.~2 we show the CEDMs for the down and strange quarks in the
SUSY SU(5) GUT with the right-handed neutrinos. We assume the minimal
supergravity scenario and take $M_{H_c}=2\times 10^{16}$GeV. 
The CP violating phases are taken to be maximal. In
Fig.~2(a) we show the strange quark CEDM as a function of the
right-handed tau neutrino mass. We take $m_{\nu_\tau}=0.05$eV and
$U_{\mu3}=1/\sqrt{2}$ and use Eq.~(\ref{seesaweq}) in order to fix the
neutrino Yukawa coupling constants.  For the SUSY breaking parameters,
we take $m_0=500$GeV, $A_0=0$, $m_{\tilde{g}}=500$GeV and
$\tan\beta=10$, which lead to $\overline{m}_{\tilde{q}}\simeq
640$GeV. We ignore the contribution from the electron and muon
neutrino Yukawa interactions to the flavor violation in the
right-handed down-type squark mass matrix. The contribution is bounded
by the constraints from the $K^0$--$\overline{K}^0$ mixing and
$Br(\mu\rightarrow e\gamma)$ when $|U_{e2}|\sim 1/\sqrt{2}$.  From
this figure, the right-handed tau neutrino mass should be smaller than
$\sim 3\times 10^{14}$GeV.  In Fig.~2(b) we show the down quark CEDM
as a function of the right-handed tau neutrino mass. This comes from
non-vanishing $U_{e3}$ in our assumption that the right-handed
neutrino mass matrix is diagonal. The current bound is not significant
even when $U_{e3}=0.2$.

The new technique for the measurement of the deuteron EDM has a great
impact on the quark CEDMs if it is realized
\cite{Semertzidis:2003iq}. If they establish the sensitivity of $d_D
\sim 10^{-27}e\,cm$, we may probe the new physics to the level of
$e d^C_s \sim 10^{-26}\;e\;cm$ and $e d^C_d \sim e d^C_u \sim
10^{-28}\;e\;cm$, which are much stronger than the bounds from the
neutron and $^{199}$Hg atom EDMs
\cite{HS2}. This may imply that we can probe the structure in the
neutrino sector even if $M_{N_3}\sim 10^{13}$GeV and  $U_{e3} \sim
0.02$.

In the above discussion we have not discussed the up quark CEDM. The
right-handed up-type squarks also have the flavor-violating mass
terms, which depend on the GUT CP-violating phases, and the magnitudes
are controlled by the CKM matrix at the GUT scale. (See
Eq.~(\ref{sfermionmass}).)  They also contribute to the hadronic EDMs
in the SUSY SU(5) GUT
\cite{Romanino:1996cn}.  However, the off-diagonal terms in both the
left-handed and right-handed up-type squark mass matrices are induced
by the bottom quark Yukawa coupling, and the up quark CEDM and EDM
induced by them are proportional to $\tan^4\beta$. We find that the
CEDM for the up quark can reach to $10^{-28}cm$ when $\tan\beta \simeq
35$ and $m_{SUSY}\simeq 500$GeV. Thus, if we observe the non-vanishing
deuteron EDM larger than $10^{-27}e cm$, it might be interpreted as
the contribution of the down or strange quark CEDM in the SUSY GUT
with the right-handed neutrinos.

When the non-vanishing hadronic EDMs are observed, it would be
important to take correlation among various processes.  One of them is
the correlation among various hadronic EDMs, including neutron and
atoms. The strange quark CEDM contribution to the neutron EDM is not
suppressed compared with other light quark ones, since it comes from
loop diagrams. On the other hand, the contribution to the EDMs for
diamagnetic atoms, such as $^{199}$Hg atom, is suppressed by the
strange quark mass. It might be possible to discriminate the
CP-violating source by comparing the various hadronic EDMs.

It is also essential to survey the correlations between the hadronic
EDMs and the lepton-flavor violating processes, which are imposed by
the GUT relation in Eq.~(\ref{gutrelation}).  The first one is the
correlation between the strange quark CEDM and
$Br(\tau\rightarrow\mu\gamma)$. In Fig.~3(a),
$Br(\tau\rightarrow\mu\gamma)$ is shown as a function of the
right-handed tau neutrino mass in Fig.~3(a).  Here, the input
parameters are the same as in Fig.~2.  Compared with Fig.~2(a), it is
found that $Br(\tau\rightarrow\mu\gamma)$ should be smaller than
$10^{-(7-8)}$.  It is argued that the super $B$ factory may reach to
$\sim 10^{-8}$ \cite{superb}. The second one is the correlation
between the down quark CEDM and $Br(\mu\rightarrow e\gamma)$. The
current bound on $Br(\mu\rightarrow e\gamma)$ already gives a
constraint on $M_{N_2}\sim 10^{13}$GeV assuming $U_{e2}\simeq
1/\sqrt{2}$.  The bound is expected to be improved by about one order
of magnitude in the MEG experiment \cite{meg}, which may reach to
$Br(\mu\rightarrow e\gamma)\sim 10^{-14}$.  In addition to it,
$Br(\mu\rightarrow e\gamma)$ is also sensitive to a product
$U_{e3}U_{\mu3}^\star$  in our assumption that the right-handed
neutrino mass matrix is diagonal. In Fig.~3(b), $Br(\mu\rightarrow e\gamma)$ is
shown as a function of the right-handed tau neutrino mass. Here, we
neglect the contribution proportional to $U_{e2}$. When $U_{\mu3}$ is
maximal, the process gives a stronger constraint on the model
parameters than the hadronic EDMs.  Since $\mu\rightarrow e\gamma$
gives an information about the structure in the neutrino sector, which
is independent of those from the hadronic EDMs, the LFV experiment is
complementary to the measurement of the deuteron EDM in the SUSY GUTs.

Finally, we discuss the hadronic EDMs in the case when the SUSY
breaking terms are generated below the GUT scale. In the above
discussion they are assumed to be generated above the GUT scale.
Below the colored Higgs masses, the interaction contributing to the
right-handed down-type squark mass terms is suppressed by $M_{H_c}^2$
as
\begin{eqnarray}
\int d^4\theta 
{\rm e}^{i\varphi_{d_{ij}}}~  U_{ik}U^\star_{jk} 
\frac{f_{\nu_k}^2}{M_{H_c}^2} \overline{N}_k^\dagger \overline{N}_k 
\overline{D}_i^\dagger \overline{D}_j
\label{dim6}
\end{eqnarray}
where $\overline{D}$ is for the chiral multiplet of the right-handed
down-type quarks. When the SUSY breaking mass terms in the MSSM are
generated by the dynamics at $M_{mess}$, which is between $M_{H_c}$
and $M_{N}$, the off-diagonal components in the right-handed squark
mass matrix generated by the interaction (\ref{dim6}) is suppressed by only
$M_{mess}^2/M_{H_c}^2$, compared with Eq.~(\ref{sfermionmass}). Thus,
when $M_{mess}$ is not much smaller than $M_{H_c}$, the sizable
off-diagonal components might be generated.

In the anomaly mediation scenario \cite{anomaly}, the SUSY breaking
terms in the MSSM are insensitive to the high energy physics, and they
are determined by the interactions and particle contents at low
energy. However, the slepton mass squared is negative in the original
model. One of the solutions is additional contribution from the
U(1)$_{B-L}$ $D$ term to the slepton mass squared.  In this extension
the UV insensitivity is broken by the right-handed neutrino mass terms
\cite{b-l}, which violates the U(1)$_{B-L}$ symmetry, even if quarks
and leptons in the MSSM do not contribute to the anomaly of
U(1)$_{B-L}$.  In this case the colored Higgs interaction generates
the off-diagonal mass terms for the right-handed down-type squarks as
\begin{eqnarray}
(m_{\tilde{d}_R}^2)_{ij}  &\simeq&-
{\rm e}^{i\varphi_{d_{ij}}}  U^\star_{ik}U_{jk} 
\frac{f_{\nu_k}^2}{(4\pi)^2} 
\frac{M_{N_k}^2}{M_{H_c}^2}
\nonumber\\&
&\times
\left\{
(2m_{\tilde{\nu}_R}^2-m_{H_c}^2-m_{\overline{H}_c}^2)
+(3m_{\tilde{\nu}_R}^2+m_{\tilde{d}_R}^2-m_{\overline{H}_c}^2)
\log\frac{M_{N_k}^2}{M_{H_c}^2}
\right\},
\label{anomaly1}
\end{eqnarray}
where $m_{H_c}$, $m_{\overline{H}_c}$ and $m_{\tilde{\nu}_R}$
are the SUSY breaking masses for the colored Higgs multiplets
and the right-handed neutrino, respectively. Since the U(1)$_{B-L}$
contribution to the SUSY breaking scalar mass squared is proportional
to the U(1)$_{B-L}$ charge $q$ as $-q\langle D_{B-L}\rangle$,
Eq.~(\ref{anomaly1}) becomes
\begin{eqnarray}
(m_{\tilde{d}_R}^2)_{ij}  &\simeq&
2 {\rm e}^{i\varphi_{d_{ij}}}  U^\star_{ik}U_{jk} 
\frac{f_{\nu_k}^2}{(4\pi)^2} 
q_N\langle D_{B-L}\rangle
\frac{M_{N_k}^2}{M_{H_c}^2}
(1+\log\frac{M_{N_k}^2}{M_{H_c}^2})
\nonumber\\
&=&
- \int d^4\theta\;
 {\rm e}^{i\varphi_{d_{ij}}}  U^\star_{ik}U_{jk} 
\frac{f_{\nu_k}^2}{(4\pi)^2} 
\frac{{\rm e}^{-2q_N V_{B-L}}M_{N_k}^2}{M_{H_c}^2}
\log\frac{{\rm e}^{-2q_NV_{B-L}}M_{N_k}^2}{M_{H_c}^2},
\label{anomaly2}
\end{eqnarray}
where $V_{B-L}=\theta^2\bar{\theta}^2 D_{B-L}$ and $q_N$ is the
U(1)$_{B-L}$ charge of the right-handed neutrinos.  If the colored
Higgs mass is lighter than the typical GUT scale $(\sim 2\times
10^{16}$GeV) and $M_{N_3}/M_{H_c}$ is not extremely small, the 
correction might be observable.

In summary, we show the hadronic EDM constraints on the neutrino
sector in the SUSY SU(5) GUT with the right-handed neutrinos. When the
CP violating phases are maximal, the hadronic EDMs are sensitive to
the right-handed down-type squark mixings, especially between the
second and third generations and between the first and third ones,
which are induced by the neutrino Yukawa interaction.  The current
experimental bound of the neutron EDM implies that the right-handed
tau neutrino mass is smaller than about $10^{14}$ GeV in the minimal
supergravity scenario, and it may be improved furthermore in future
experiments, such as in measurement of the deuteron EDM. When the
non-vanishing hadronic EDMs are observed, we can probe the structure
in the SUSY GUT models by investigating the correlations between the
results of the hadronic EDM and LFV searches.

\section*{Acknowledgments}
The work of J.H. is supported in part by the Grant-in-Aid for Science
Research, Ministry of Education, Science and Culture, Japan
(No.15540255, No.13135207 and No.14046225).  That of Y.S. is
supported in part by the 21st century COE program, ``Exploring New
Science by Bridging Particle-Matter Hierarchy''. Also, that of M.K.
is supported in part by JSPS.

\newpage
%
%
\newcommand{\Journal}[4]{{\sl #1} {\bf #2} {(#3)} {#4}}
\newcommand{\APJ}{Ap. J.}
\newcommand{\CJP}{Can. J. Phys.}
\newcommand{\MPL}{Mod. Phys. Lett.}
\newcommand{\NC}{Nuovo Cimento}
\newcommand{\NP}{Nucl. Phys.}
\newcommand{\PL}{Phys. Lett.}
\newcommand{\PR}{Phys. Rev.}
\newcommand{\PRep}{Phys. Rep.}
\newcommand{\PRL}{Phys. Rev. Lett.}
\newcommand{\PTP}{Prog. Theor. Phys.}
\newcommand{\SJNP}{Sov. J. Nucl. Phys.}
\newcommand{\ZP}{Z. Phys.}

\newpage

\begin{table}
\begin{center}
\begin{tabular}{|c|c||c|c|}
\hline
\multicolumn{4}{|c|}
{$^{199}$Hg EDM (neutron EDM) ($m_{SUSY}=500$GeV and $\tan\beta=10$)}
\\ \hline
$(\delta^u_{12})_{R}(\delta^u_{21})_{L}$                       
&  
$0.8(1)\times 10^{-3}$ &
$(\delta^u_{13})_{R}(\delta^u_{31})_{L}$ 
&  
$3(5)\times 10^{-6}$ 
\\ \hline
$(\delta^d_{12})_{R}(\delta^d_{21})_{L}$                       
&  
$0.6(1)\times 10^{-3}$ &
$(\delta^d_{13})_{R}(\delta^d_{31})_{L}$ 
&  
$2(4)\times 10^{-5}$ 
\\ \hline
$(\delta^d_{23})_{R}(\delta^d_{32})_{L}$                       
&  
$3(0.2)\times 10^{-3}$ &
&  
\\ 
\hline\hline
\multicolumn{4}{|c|}
{$\Delta M_K$  ($m_{SUSY}=500$GeV)}
\\ \hline
$(\delta^d_{12})_{R}^2$                       
&  
$2\times 10^{-3}$ &
$(\delta^d_{13})_{R}^2(\delta^d_{32})^2_{R}$ 
&  
$3\times 10^{-2}$ 
\\ \hline
$(\delta^d_{12})_{R}(\delta^d_{12})_{L}$                       
&  
$7\times 10^{-6}$
&
$(\delta^d_{13})_{R}(\delta^d_{32})_{R}(\delta^d_{13})_{L}(\delta^d_{32})_{L}$                       
&  
$2\times 10^{-5}$
\\ \hline
$(\delta^d_{12})_{R}(\delta^d_{13})_{R}(\delta^d_{32})_{R}$ 
&  
$3\times 10^{-3}$
&
$(\delta^d_{12})_{R}(\delta^d_{13})_{L}(\delta^d_{32})_{L}$  
&  
$1\times 10^{-5}$
\\
\hline\hline
\multicolumn{4}{|c|}
{$\epsilon_K$ ($m_{SUSY}=500$GeV)}
\\ \hline
$(\delta^d_{12})_{R}^2$                       
&  
$1\times 10^{-5}$ &
$(\delta^d_{13})_{R}^2(\delta^d_{32})^2_{R}$ 
&  
$2\times 10^{-4}$ 
\\ \hline
$(\delta^d_{12})_{R}(\delta^d_{12})_{L}$                       
&  
$5\times 10^{-8}$
&
$(\delta^d_{13})_{R}(\delta^d_{32})_{R}(\delta^d_{13})_{L}(\delta^d_{32})_{L}$                       
&  
$1\times 10^{-7}$
\\ \hline
$(\delta^d_{12})_{R}(\delta^d_{13})_{R}(\delta^d_{32})_{R}$ 
&  
$2\times 10^{-5}$
&
$(\delta^d_{12})_{R}(\delta^d_{13})_{L}(\delta^d_{32})_{L}$  
&  
$7\times 10^{-8}$
\\ 
\hline\hline
\multicolumn{4}{|c|}
{$\Delta M_D$ ($m_{SUSY}=500$GeV)}
\\ \hline
$(\delta^u_{12})_{R}^2$                       
&  
$1\times 10^{-2}$ &
$(\delta^u_{13})_{R}^2(\delta^u_{32})^2_{R}$ 
&  
$2\times 10^{-1}$ 
\\ \hline
$(\delta^u_{12})_{R}(\delta^u_{12})_{L}$                       
&  
$3\times 10^{-4}$
&
$(\delta^u_{13})_{R}(\delta^d_{32})_{R}(\delta^u_{13})_{L}(\delta^u_{32})_{L}$                       
&  
$6\times 10^{-4}$
\\ \hline
$(\delta^u_{12})_{R}(\delta^u_{13})_{R}(\delta^u_{32})_{R}$ 
&  
$2\times 10^{-2}$
&
$(\delta^u_{12})_{R}(\delta^u_{13})_{L}(\delta^u_{32})_{L}$  
&  
$4\times 10^{-4}$
\\ 
\hline\hline
\multicolumn{4}{|c|}
{$\Delta M_B$ ($m_{SUSY}=500$GeV)}
\\ \hline
$(\delta^d_{13})_{R}^2$                       
&  
$1\times 10^{-2}$ &
$(\delta^d_{13})_{R}(\delta^d_{13})_{L}$                       
&  
$3\times 10^{-4}$
\\ 
\hline\hline
\multicolumn{4}{|c|}
{$Br(b\rightarrow s\gamma$) ($m_{SUSY}=500$GeV and $\tan\beta=10$)}
\\ \hline
$(\delta^d_{23})_{R}$                       
&  
$0.6$ &
&  
\\
\hline
\end{tabular}
\end{center}
\label{table1}
\caption{Constraints on the mass insertion parameters of squarks from 
neutral $K$, $D$ and $B$ meson mixings, $b\rightarrow s\gamma$, and
EDMs of neutron and $^{199}$Hg atom. Here, we evaluate the gluino diagram
contribution to them and require them to be smaller than the
experimental values or the bounds as $\Delta m_K<3.5\times
10^{-12}$MeV, $ \epsilon_K<2.3\times 10^{-3}$, $\Delta m_D<1.3\times
10^{-10}$MeV, $\Delta m_B<3.8\times 10^{-10}$MeV, and $Br(b\rightarrow
s\gamma)<3.3\times 10^{-4}$. For the hadronic EDMs, see text. We take
the SUSY particle masses equal to $m_{SUSY}$.  Constraints on the
combination of the mass insertion parameters in this table are
proportional to $m_{SUSY}^2$. Also, we consider only diagrams
proportional to $\tan\beta$ for the CEDMs of the strange and down
quarks and $b\rightarrow s\gamma$.  }
\end{table}

\begin{table}
\begin{center}
\begin{tabular}{|c|c||c|c|}
\hline
\multicolumn{4}{|c|}
{$Br(l\rightarrow l'\gamma)$  ($m_{SUSY}=200$GeV and $\tan\beta=10$)}
\\ \hline
$(\delta^l_{12})_{L}$                       
&  
$2\times 10^{-4}$ &
$(\delta^l_{13})_{L}(\delta^l_{32})_{L}$ 
&  
$4\times 10^{-4}$ 
\\ \hline
$(\delta^l_{12})_{R}$                       
&  
$3\times 10^{-3}$ &
$(\delta^l_{13})_{R}(\delta^l_{32})_{R}$ 
&  
$3\times 10^{-3}$ 
\\ \hline
$(\delta^l_{13})_{L}(\delta^l_{32})_{R}$                       
&  
$1\times 10^{-4}$ 
&
$(\delta^l_{13})_{R}(\delta^l_{32})_{L}$                       
&  
$1\times 10^{-4}$ 
\\ \hline
$(\delta^l_{23})_{L}$
&  
$8\times 10^{-2}$ 
&
$(\delta^l_{13})_{L}$                       
&  
$8\times 10^{-2}$ 
\\ \hline
$(\delta^l_{23})_{R}$
&  
$1$ 
&
$(\delta^l_{13})_{R}$                       
&  
$1$
\\ 
\hline\hline
\multicolumn{4}{|c|}
{Electron EDM  ($m_{SUSY}=200$GeV and $\tan\beta=10$)}
\\ \hline
$(\delta^l_{12})_{R}(\delta^l_{21})_{L}$                       
&  
$3\times 10^{-4}$ &
$(\delta^l_{13})_{R}(\delta^l_{31})_{L}$ 
&  
$2\times 10^{-5}$ 
\\ \hline
\hline
\end{tabular}
\end{center}
\label{table2}
\caption{
Constraints on the mass insertion parameters of sleptons from
$Br(\mu\rightarrow e \gamma)$, $Br(\tau\rightarrow \mu(e) \gamma)$ and
the electron EDMs. Here, we take the SUSY particle masses equal to
$m_{SUSY}$ and consider the contribution to them proportional to
$\tan\beta$. The experimental bounds on the processes are
$Br(\mu\rightarrow e \gamma)<1.2\times 10^{-11}$, $Br(\tau\rightarrow
\mu(e) \gamma)<3.6(3.2)\times 10^{-7}$, and $d_e<4.3\times 10^{-27}e~cm$.}
\end{table}

\begin{figure}[p]
\label{fig:CEDM}
\begin{center} 
\begin{picture}(455,140)(30,-20)
\SetOffset(100,0)
\ArrowArcn(135,25)(75,180,90)
\ArrowArc(135,25)(75,0,90)
\Vertex(135,100){3}
\Text(135,120)[]{$\tilde{g}$}

\ArrowLine(60,25)(30,25)
\DashArrowLine(90,25)(60,25){3}             \Vertex(90,25){3}
\DashArrowLine(135,25)(90,25){3}             \Vertex(135,25){3}
\DashArrowLine(135,25)(180,25){3}       \Vertex(185,25){3}  
\DashArrowLine(180,25)(210,25){3}        
\ArrowLine(210,25)(240,25)

\Text(45,15)[]{$q_{Li}$}
\Text(70,15)[]{$\tilde{q}_{Li}$}
\Text(115,15)[]{$\tilde{q}_{Lj}$}
\Text(155,15)[]{$\tilde{q}_{Rj}$}
\Text(200,15)[]{$\tilde{q}_{Ri}$}
\Text(225,15)[]{$q_{Ri}$}

\Photon(172,110)(195,135){2}{5}
\Text(203,145)[]{$g$}

\end{picture} 

\caption{ Dominant diagram contributing to the CEDMs of light
quarks when both the left-handed and right-handed squarks have 
flavor mixings.
}

\end{center}
\end{figure}
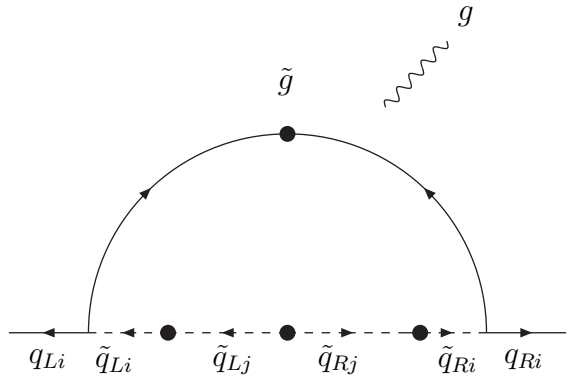

\begin{figure}
 \centerline{
\epsfxsize = 0.5\textwidth \epsffile{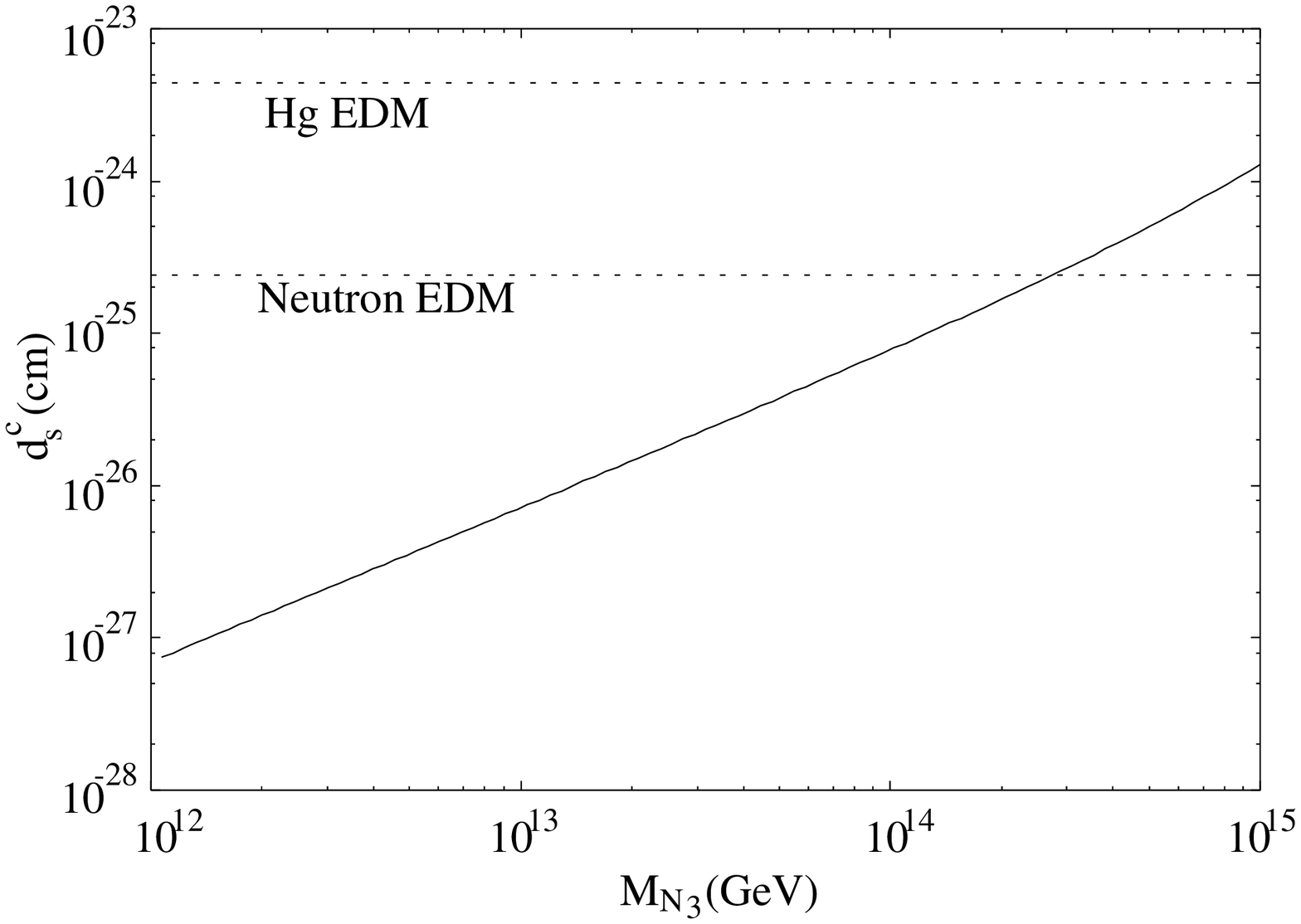} 
\epsfxsize = 0.5\textwidth \epsffile{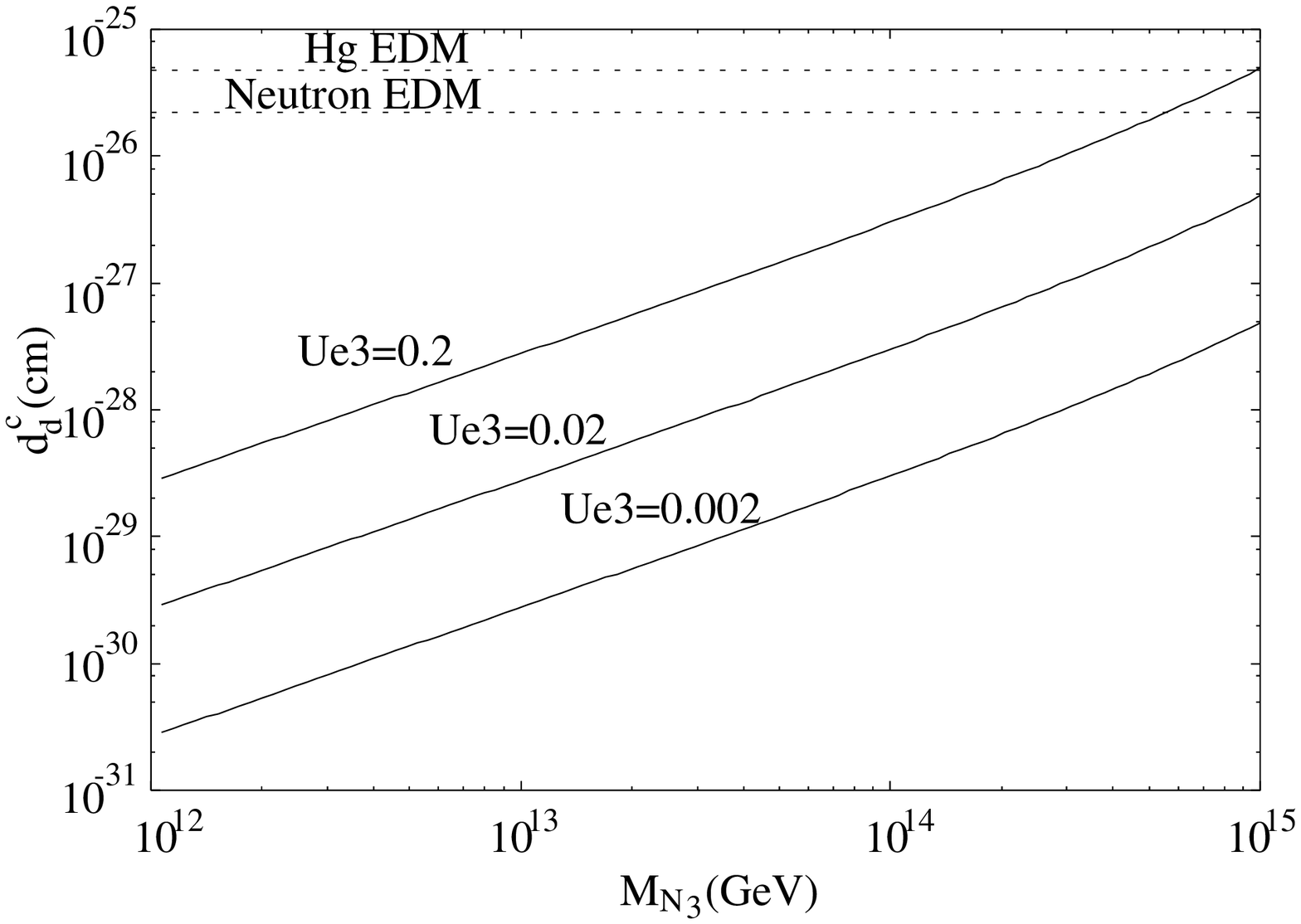} 
}
\vspace*{-5mm}
\caption{
CEDMs for the strange quark in (a) and for the down quark in (b) as
functions of the right-handed tau neutrino mass, $M_{N_3}$.  Here,
$M_{H_c}=2\times 10^{16}$GeV, $m_{\nu_\tau}=0.05$eV,
$U_{\mu3}=1/\sqrt{2}$, and $U_{e3}=0.2$, 0.02, and 0.002.  For the MSSM
parameters, we take $m_0=500$GeV, $A_0=0$, $m_{\tilde{g}}=500$GeV and
$\tan\beta=10$.}
\end{figure}

\begin{figure}
 \centerline{
\epsfxsize = 0.5\textwidth \epsffile{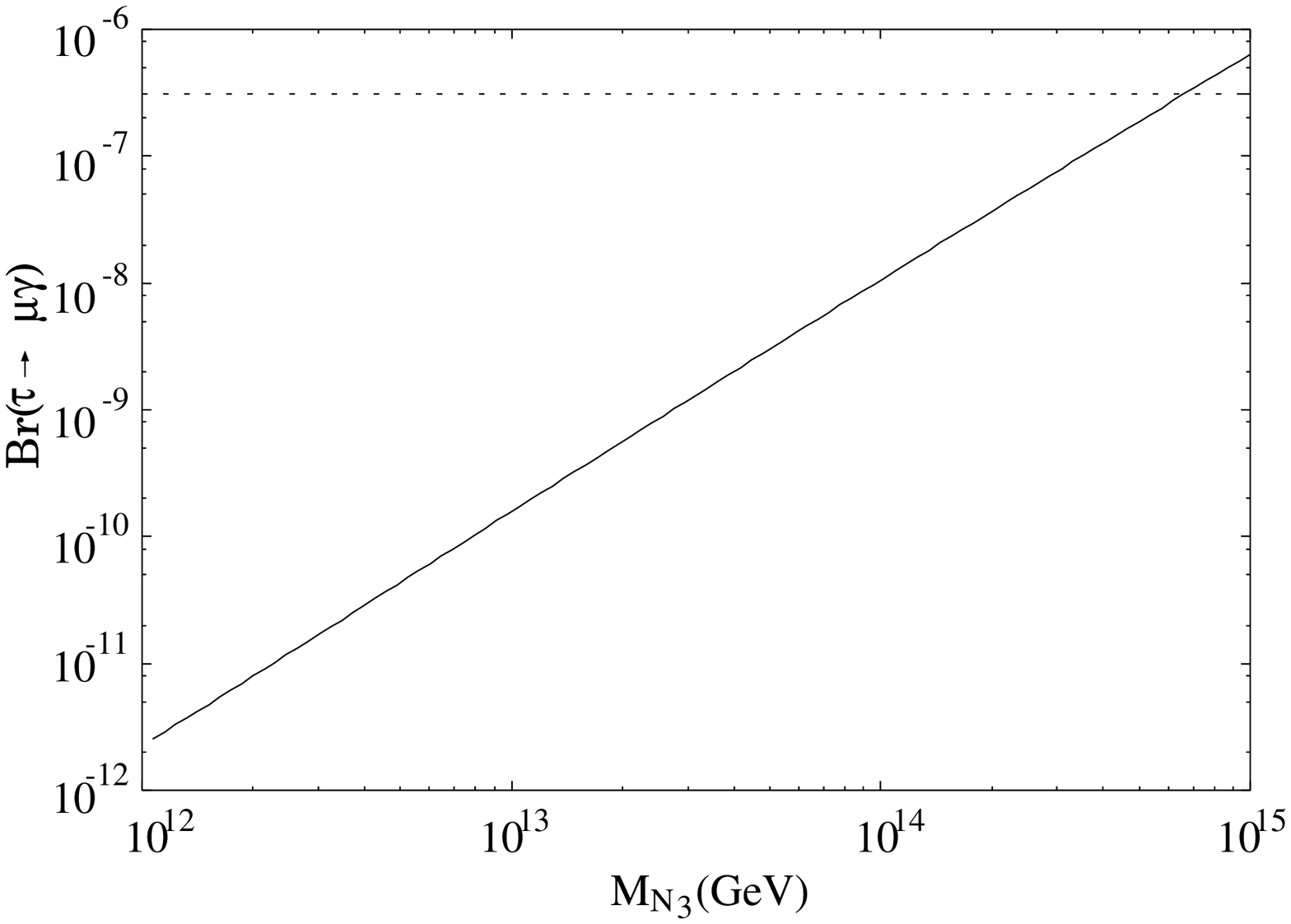} 
\epsfxsize = 0.5\textwidth \epsffile{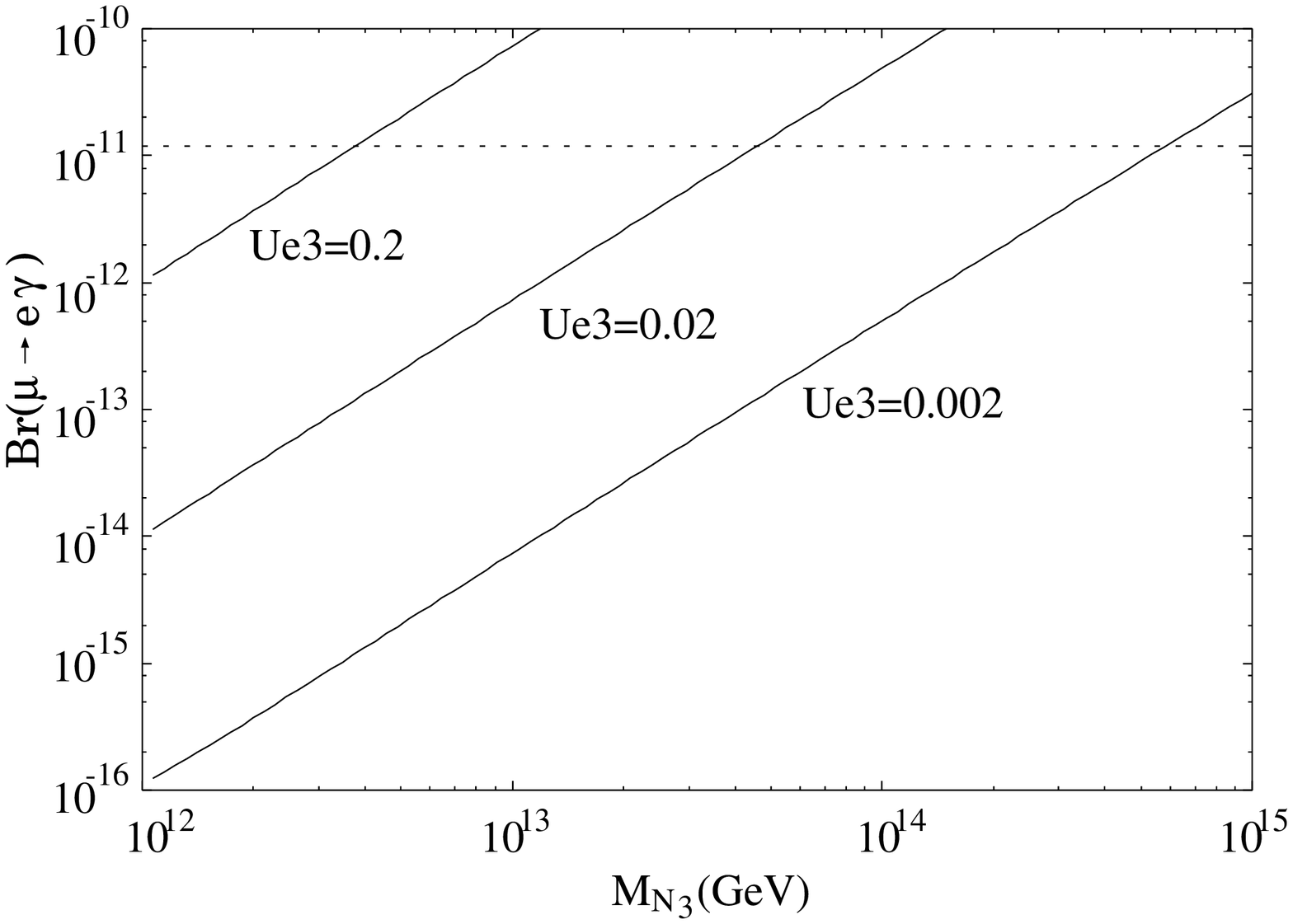} 
}
\vspace*{-5mm}
\caption{
$Br(\tau\rightarrow\mu\gamma)$ in (a) and $Br(\mu\rightarrow e\gamma)$ (b) as
functions of the right-handed tau neutrino mass, $M_{N_3}$.  Here,
the input parameters are the same as in Fig.~2. The dashed lines are
the experimental bounds \cite{Abe:2003sx,Brooks:1999pu}. }
\end{figure}

\end{document}